\documentclass[twoside,twocolumn,pra]{revtex4-2}
\usepackage{graphicx}
\usepackage{dcolumn}
\usepackage{bm}
\usepackage{amsmath}
\usepackage{stmaryrd}
\usepackage[english]{babel}
\usepackage{tcolorbox}
\UseRawInputEncoding
\usepackage{lipsum}
\usepackage{subcaption}   
\usepackage{float}
\usepackage{graphics} 

\begin{document}
\title{Quantum Probability from Temporal Structure}
\author{Michael Ridley}
\email{ridley@tau.ac.il}
\affiliation{School of Physics and Astronomy, Tel-Aviv University, Tel-Aviv 69978, Israel}

\begin{abstract}
The Born probability measure describes the statistics of measurements in which observers self-locate themselves in some region of reality. In $\psi$-ontic quantum theories, reality is directly represented by the wavefunction. We show that quantum probabilities may be identified with fractions of a universal multiple-time wavefunction containing both causal and retrocausal temporal parts. This wavefunction is defined in an appropriately generalized history space on the Keldysh time contour. Our deterministic formulation of quantum mechanics replaces the initial condition of standard Schr{\"o}dinger dynamics with a network of `fixed points' defining quantum histories on the contour. The Born measure is derived by summing up the wavefunction along these histories. We then apply the same technique to the derivation of the statistics of measurements with pre- and post-selection.\end{abstract}

\maketitle

\section{Introduction} Textbook formulations of quantum mechanics contain 

(a) an \emph{ontological} postulate - the state of a physical system is represented by a wavefunction $\left|\Psi\right\rangle$.

(b) a \emph{dynamical} postulate - the state evolves deterministically according to the time-dependent Schr{\"o}dinger equation (TDSE).

(c) a \emph{composition} postulate - the state space of a composite system is the tensor product of the spaces of its subsystems.

(d) a \emph{statistical} postulate - the probability of each measurement outcome is given by the Born measure. 

Explaining the appearance of a probabilistic element in postulate (d) is perhaps the main obstacle to understanding quantum theory \cite{feynman_concept_1951}. In $\psi$-ontic quantum theories, the universal wavefunction is in direct correspondence with physical reality \cite{pusey_reality_2012,colbeck_is_2012,ringbauer_measurements_2015}. There have been many attempts to derive and/or explain (d) from postulates (a-c) within a $\psi$-ontic framework \cite{deutsch_quantum_1999,zurek_environment-assisted_2003,zurek_probabilities_2005,wallace_emergent_2012,sebens_self-locating_2018} with no reference to the physical `collapse' of wavepackets, and thereby solve the hard part of the measurement problem. Some derive the Born measure by placing `rationality' constraints on the beliefs of observers \cite{deutsch_quantum_1999,wallace_emergent_2012}, but such theories have it backwards - rational beliefs do not determine regularities in nature. Rather, the structure of nature grounds measurement statistics, and therefore determines what it is rational to believe. Other derivations use symmetry arguments \cite{zurek_environment-assisted_2003,zurek_probabilities_2005,sebens_self-locating_2018}, but these rely on auxiliary formal assumptions and a separation of the quantum state into the system plus the environment. Also, these approaches are based upon vigorously-debated concepts of probability, rather than the ontology of the physical theory itself. A new perspective on the problem of probability in quantum mechanics is sorely needed. 

The concept of the probability of self-location avoids the need for randomness in quantum mechanics, and enables the assignment of probabilities to branches of the wavefunction without recourse to genuine randomness in nature \cite{vaidman_all_2016,sebens_self-locating_2018,vaidman_derivations_2020}. However, in the present work we go further: since the self-locating of an observer is carried out with respect to the wavefunction itself, the probability of being located in a region of the wavefunction should be literally equated with its relative proportion of the total wavefunction, and therefore grounded in physical ontology alone, i.e. \textit{the probability measure must therefore emerge from the internal structure of the wavefunction itself}. This invites a $\psi$-ontic explanation of the appearance of chance in quantum mechanics: an observer is localized to a region of the wavefunction consistent with experiment. The Born rule quantifies the relative amount of reality, or the ‘measure of existence’ of that region \cite{vaidman_schizophrenic_1998,mandolesi_quantum_2020,vaidman_derivations_2020}.

The problem of treating time quantum mechanically is an apparently unrelated foundational question, which however has recently attracted a great deal of attention. In particular, time appears as a background parameter in postulate (b), but this way of representing time is at odds with the geometric notion of time in general relativity \cite{horwitz_two_1988,maccone_fundamental_2019}. Proposals for measurements of quantum time generally focus on the theoretical absolute time of a quantum state, and
the introduction of a background `quantum clock' degree of freedom to measure `arrival times' of particles at a detector arising from entanglement between subsystems
\cite{page_evolution_1983,marletto_evolution_2017,maccone_quantum_2020}. Traditionally, the problem of defining time in quantum mechanics is presented
as the problem of defining an Hermitian operator with monotonically increasing eigenvalues for a system with Hamiltonian bounded from below \cite{pauli_allgemeinen_1933}. It can be shown
that such an operator always leads to finite amplitudes for the reverse-time process \cite{unruh_time_1989}. 

Far from being a hindrance to the description of quantum time however, we may elevate reverse-time causal processes to a central feature of the theory implicit in the unitary evolution of states with complex amplitudes. In 1964 \cite{aharonov_time_1964}, Aharonov \emph{et al.} published the time-symmetric two-state vector formalism (TSVF) \cite{aharonov_two-state_2008,aharonov_two-time_2017}, describing the probabilities of measurements sandwiched between pre- and post-selections with the Aharonov-Bergmann-Lebowitz (ABL) rule. The TSVF was later generalised to a multiple-time formalism assigning a Hilbert space $\mathcal{H}_{t}$ and its conjugate space $\mathcal{H}^{\dagger}_{t}$ (for backwards-directed states) to each instant of time, i.e. the composition postulate (c) was applied to treat time instants as distinct quantum subsystems \cite{aharonov_is_1984,aharonov_multiple-time_2009}. The wavefunction is then a global time-extended structure composed of \textit{temporal parts} \cite{heller_temporal_1984}. It was recently shown that this assignment of two Hilbert spaces to each moment of time is necessary to capture all the correlations in the quantum dynamical evolution of a particle with an equivalent multipartite state \cite{aharonov_each_2014}. The experimental success of the TSVF \cite{lundeen_experimental_2009,vaidman_past_2013,curic_experimental_2018}, various explicitly time symmetric formulations \cite{watanabe_symmetry_1955,cramer_transactional_1986,wharton_time-symmetric_2007} and recent demonstrations of indefinite causal ordering \cite{leifer_is_2017,zych_bells_2019, castro-ruiz_quantum_2020,drummond_retrocausal_2020} all provide evidence for a more complex causal structure in nature than a single background time parameter can offer.  

By coincidence, the year 1964 saw publication of another time-symmetric formalism by Keldysh \cite{keldysh_diagram_1964}. The resulting Nonequilibrium Green's function (NEGF) theory describes the propagation of correlation functions along a time contour $C$ composed of both forwards ($f$) and backwards ($b$) time branches \cite{tang_full-counting_2014,esposito_quantum_2015,aeberhard_microscopic_2017,chaudhuri_probing_2019,schlunzen_achieving_2020,tuovinen_comparing_2020,atanasova_correlated_2020}. Note that the contour time structure itself does not logically presuppose the Born measure, although propagating statistical averages on this contour is equivalent to weighting them with Born probabilities.

In this paper we take advantage of this logical equivalence, showing that the derivation of the Born measure is possible from unitary dynamics and wavefunction structure alone given a wavefunction-based definition of probability. We incorporate the full Keldysh causal structure of quantum mechanics within the universal wavefunction, and model temporally local events in terms of `fixed point' boundary conditions. We therefore refer to this version of quantum mechanics as the fixed point formulation (FPF). We then introduce a statistical postulate based on our probability definition and derive the correct probability measure from ontological and dynamical postulates describing unitary evolution in Hilbert space without random collapse. Thus, we reduce the number of independent postulates in the quantum theory - the Born measure follows from ontology, composition and dynamics.

\section{The universal wavefunction}
\subsection{General considerations}
We wish to focus on the global temporal structure of wavefunctions composed of both macroscopic and microscopic parts, without approximation or tracing out environmental degrees of freedom. Such a wavefunction represents the observer, the system being observed and the environment in a typical quantum experiment, and it is in this sense that we refer to it as `universal'. We start from a strong $\psi$-ontic standpoint, with the following conceptual desiderata:
\begin{itemize}
    \item[] \textbf{Completeness}:
    The wavefunction is all that exists - it contains all physical properties of nature at all moments of time, which exist eternally. 
    \item[] \textbf{Measurement Physicality}:
    Measurements are physical processes occurring within temporal regions of the universal wavefunction.
    \item[] \textbf{Event Symmetry}: The local description of nature is independent of event location. There are no ontologically privileged spacetime points.
    \item[] \textbf{Self-Location}:
    Temporal boundary constraints provide the only information an observer can use to locate themselves within the wavefunction.
\end{itemize}
  
The concept of probability developed here utilises the principle of \textbf{Self-Location}:

\textbf{Definition 1} (Quantum probability)

\emph{In a temporal region of the wavefunction defined by some set of constraints, process $A$ has probability $\textrm{p}(A)=x$ if and only if $A$ occurs in a fraction $x$ of the total available wavefunction.}

Given said constraints, an observer should set their subjective degree of belief that they are located in a region of reality where $A$ occurs to the corresponding fraction of reality, i.e. to the quantum probability. This approach is \textit{logically minimal, physically maximal} - it grounds the mathematical theory of probability in physical ontology. A proponent of \textbf{Completeness} must then answer the question:

\textit{Which structural feature of the wavefunction implies the Born measure?}

To begin to answer this, we observe that recent works in quantum cosmology describe physical systems with sequences of time-indexed properties (described by projection operators), or `histories' \cite{griffiths_what_2017,hartle_one_2017}. Given a time ordering of $N_t$ times at which physical properties are instantiated, $t_{N_t}>t_{N_{t}-1}>\ldots>t_{1}$, reality can be described by a `universal' wavefunction $\left|\Psi_{U}\right\rangle$ specifying the full set of histories defined on these times. In the histories formalism, each value of the time $t_{i}$ labels a distinct subspace $\mathcal{H}_{t_{i}}$ of the \textit{history} Hilbert space \cite{isham_continuous_1995,isham_continuous_1998}:
\begin{equation}
  \mathcal{H}_{H}\equiv\mathcal{H}_{t_{N_{t}}}\otimes\ldots\otimes\mathcal{H}_{t_{1}}
  \label{eqn:Hilbert_Space_forwards}
\end{equation}  

In this space, history states can be viewed as 'records' of all the different stages in a quantum process, indexed by time. Thus the states at distinct times enter the wavefunction in an atemporal fashion suited to a block universe point of view.

Parallel to the consistent histories approach, products of time-localized Hilbert spaces feature in the time-symmetric approach to quantum mechanics pioneered by Aharonov et al. \cite{aharonov_time_1964}. This approach, which became the two state vector formalism (TSVF) \cite{aharonov_two-state_2008} and its multiple time generalization \cite{aharonov_multiple-time_2009,aharonov_each_2014}, treats quantum measurements which include dynamical boundary conditions on past and future times symmetrically.

This is useful in the description of a system defined at time $t$ occurring between preselection and postselection measurements at the times $t_{1}$ and $t_{2}$, respectively. The preselected state $\left|\psi\left(t_{1}\right)\right\rangle$ then travels forwards in time across the interval $\left[t_{1},t\right]$ in accordance with the TDSE, and the postselected state is represented by a vector in the conjugate space $\left\langle \phi\left(t_{2}\right)\right|$ which propagates backwards across the time interval $\left[t,t_{2}\right]$. The two oppositely orientated parts of the system can then be combined into a single `two state vector'

\begin{equation}\label{eq:TSV}
    \left\langle \phi\left(t_{2}\right)\right|\otimes\left|\psi\left(t_{1}\right)\right\rangle, 
\end{equation}

which exists in the composite Hilbert space constructed from distinct time-localized `universes' existing at single times \cite{aharonov_each_2014}

\begin{equation}
    \mathcal{H}_{t_{2}}^{\dagger}\otimes\mathcal{H}_{t_{1}}
\end{equation}

States in this Hilbert space are fundamentally (i) time non-local objects and (ii) built out of parts with opposite time orientations, which immediately suggests that this is a promising avenue to explore for the development of a quantum theory of events. On this account, the solution to the apparent asymmetry under time reversal in quantum mechanics is to revise the notion of a quantum state itself to include two time degrees of freedom. 

According to the TSVF, to obtain the probability of measuring the system in some state $\left|a_{i}\right\rangle$ at the intermediate time $t \in \left[t_{1},t_{2}\right]$, the system is propagated in \emph{both} time directions, from $t_{1}\rightarrow t$ and $t_{2}\rightarrow t$, such that the amplitude of the $a_{i}$-th outcome is given by sandwiching this state between the forwards and backwards-oriented parts of Eq. \ref{eq:TSV}:

\begin{equation}
    \left\langle \phi\left(t_{2}\right)\right|U\left(t_{2},t\right)\left|a_{i}\right\rangle \left\langle a_{i}\right|U\left(t,t_{1}\right)\left|\psi\left(t_{1}\right)\right\rangle 
\end{equation}

Then, assuming the Born rule, the normalized modulus-square of this yields the probability to obtain outcome $a_{i}$:

\begin{equation}
    P_{a_{i}}=\frac{\left|\left\langle \phi\left(t_{2}\right)\right|U\left(t_{2},t\right)\left|a_{i}\right\rangle \left\langle a_{i}\right|U\left(t,t_{1}\right)\left|\psi\left(t_{1}\right)\right\rangle \right|^{2}}{\underset{k}{\sum}\left|\left\langle \phi\left(t_{2}\right)\right|U\left(t_{2},t\right)\left|a_{k}\right\rangle \left\langle a_{k}\right|U\left(t,t_{1}\right)\left|\psi\left(t_{1}\right)\right\rangle \right|^{2}}
\end{equation}

This is the ABL probability rule. In the quantum theory, it thus appears that the past and future affect each other symmetrically \cite{vaidman_past_2013}. However the TSVF relies upon a wavefunction with temporal parts whose behaviour depends on time position. Specifically, the preselected state at $t_{1}$ is a source of physical processes occurring between $t_{1}$ and $t$, and the postselected state at $t_{2}$ is a source for processes connecting $t_{2}$ to $t$. However, the state at time $t$ serves as a unique sink for both types of processes. Clearly, this is a violation of \textbf{Event Symmetry} - if, given two connected points in time, one is a source and the other a sink, and the dynamics is allowed to be time symmetric, it must follow that both points are sources and both are sinks for all time regions they are connected to.

\subsection{The universal wavefunction on the Keldysh contour}
\begin{figure}
  \includegraphics[width=.9\linewidth]{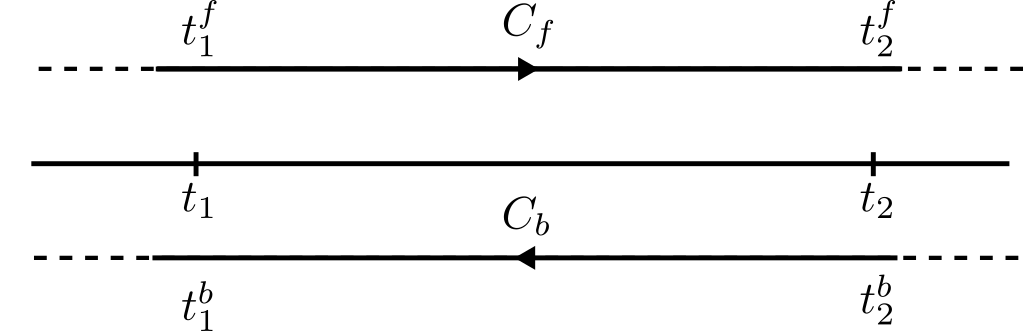}
  \caption{The Keldysh time contour in the time interval $\left[t_{1},t_{2}\right]$.}
  \label{fig:Keldysh_Contour}
\end{figure}

A wavefunction-based theory must contain a representation of the temporal processes occurring in field theories defined on an appropriate time domain. For systems consisting of particles obeying fermionic or bosonic statistics, this is done using the NEGF formalism, which is used to evaluate time-dependent expectation values of quantum observables $O\left(t_{2}\right)$ propagated from some initial time $t_{1}$:

\begin{equation}
 O\left(t_{2}\right) =  \mbox{Tr}\left[\rho_{1}U\left(t_{1},t_{2}\right)\hat{O}\left(t_{2}\right)U\left(t_{2},t_{1}\right)\right],
  \label{eqn:O_t}
\end{equation}  

where $\rho_{1}$ is the density matrix at $t_{1}$ and $U\left(t_{2},t_{1}\right)$ is the unitary evolution between times $t_{1}$ and $t_{2}$. The expression in Eq. \eqref{eqn:O_t} can be evaluated via two separate propagations, the first running forwards in time from $t_{1}$ to $t_{2}$, at which the operator $\hat{O}$ acts, before the system is propagated backwards from $t_{2}$ to $t_{1}$. This can be visualized in terms of propagation along the Keldysh time contour shown in Fig. \ref{fig:Keldysh_Contour}. The Keldysh contour consists of an `upper' branch $C_{f}$ of times $t^{f}$ on which the wavefunction travels in the forwards direction, and a `lower' branch $C_{b}$ of times $t^{b}$ on which the dynamics is reversed.

We propose a similar physical state space to $\mathcal{H}_{H}$, with the caveat that temporal degrees of freedom take values on both branches of the Keldysh time contour. Thus, given an ordering of $N_{t}$ times $t_{N_{t}}>t_{N_{t}-1}>...>t_{1}$, there are two corresponding causal orderings, one on each branch of $C \equiv C_{b} \oplus C_{f}$:
\begin{gather}\label{eqn:f_b_ordering}
t_{N_{t}}^{f}>_{C}t_{N_{t}-1}^{f}>_{C}\ldots>_{C}t_{1}^{f}\\
t_{N_{t}}^{b}<_{C}t_{N_{t}-1}^{b}<_{C}\ldots<_{C}t_{1}^{b}
\end{gather}

where the contour-ordering notation $>_{C}$, $<_{C}$ is introduced as in Ref. \cite{stefanucci_nonequilibrium_2013}. This is the main innovation of the Keldysh contour: ordering in time is distinct from causal ordering, since causal influences propagate in the antichronological direction on the lower branch $C_{b}$. 

Each of the $N_{t}$ times in a history possesses two associated Hilbert spaces for the $f$ and $b$ components. Hence, the universal wavefunction has $2N_{t}$ temporal degrees of freedom and is a member of the \textit{contour} Hilbert space:
\begin{equation}
  \mathcal{H}_{C}=\mathcal{H}_{t_{N_{t}}}^{b}\otimes\mathcal{H}_{t_{N_{t}}}^{f}\otimes\ldots\otimes\mathcal{H}_{t_{1}}^{b}\otimes\mathcal{H}_{t_{1}}^{f}
  \label{eqn:Hilbert_Space}
\end{equation}  

A wavefunction in this space is not defined at a single fixed `present', but at a sequence of moments with oppositely oriented  temporal parts acting as `source' or `sink' states for processes on the branches $C_{f}$ and $C_{b}$.

We make a corresponding first postulate:
\begin{tcolorbox}
\textbf{Ontological postulate}

\textit{The universal wavefunction $\left|\Psi_{U}\right\rangle \in \mathcal{H}_{C}$ is a `stack' of $2N_{t}$ temporal parts with fixed ordering on $C$, dividing time into $2(N_{t}-1)$ separate regions:}
\begin{equation}
\left|\Psi_{U}\right\rangle =\bigotimes_{i=1}^{N_{t}}\left|\Psi^{b}\left(t_{i}^{b}\right)\right\rangle \otimes\left|\Psi^{f}\left(t_{i}^{f}\right)\right\rangle 
\label{eqn:Psi_Universe}
\end{equation}
\end{tcolorbox}

Here, $\left|\Psi^{\alpha}\left(t_{i}^{\alpha}\right)\right\rangle$  is restricted to the $C_{\alpha}$ time branch, and in general $\left|\Psi^{f}\left(t_{i}^{f}\right)\right\rangle\neq\left|\Psi^{b}\left(t_{i}^{b}\right)\right\rangle$. The inner product is defined on the Hilbert space $\mathcal{H}_{t_{i}}^{\alpha}$ in the usual way, such that $\left\langle \Psi_{U}\right|\left.\Psi_{U}\right\rangle =1$, which implies $\left\langle \Psi^{\alpha}\left(t_{i}^{\alpha}\right)\right|\left.\Psi^{\alpha}\left(t_{i}^{\alpha}\right)\right\rangle =1$ for any $\alpha$. Oppositely-oriented parts of the wavefunction are connected independently on $C_{f}$ and $C_{b}$. We note that Eq. \eqref{eqn:Psi_Universe} can be generalized to contain a summation over all possible tensor products of time-localized states and thereby represent all multiple-time processes on the Keldysh contour, but for the purposes of the present work we focus on the case of fixed sequence size $N_t$, following the histories formulation \cite{griffiths_what_2017}. We now introduce the second core postulate:

\begin{tcolorbox}
\textbf{Dynamical postulate}

\textit{The time derivative of the wavefunction at each point on $C$ is given by the TDSE:}
\begin{equation}
i\hbar\partial_{t^{\alpha}}\left|\Psi^{\alpha}\left(t^{\alpha}\right)\right\rangle =H^{\alpha}\left(t^{\alpha}\right)\left|\Psi^{\alpha}\left(t^{\alpha}\right)\right\rangle 
\label{eqn:Branch_TDSE}
\end{equation}
\end{tcolorbox}

Note that in every case of physical interest the Hamiltonian operator is branch-independent, i.e. it takes on values on the upper/lower branches which are equal for the same physical time, $H^{b}\left(t^{b}\right)=H^{f}\left(t^{f}\right)$. For simplicity, indices on time arguments are dropped, $\left|\Psi^{\alpha}\left(t^{\alpha}\right)\right\rangle\equiv\left|\Psi^{\alpha}\left(t\right)\right\rangle$. 

The TDSE in Eq. (\ref{eqn:Branch_TDSE}) defines a unitary mapping $U^{\alpha}\left(t_{2},t_{1}\right):\mathcal{H}_{t_{1}}^{\alpha}\mapsto\mathcal{H}_{t_{2}}^{\alpha}$ between the Hilbert spaces of different times on a single branch $\left|\Psi^{\alpha}\left(t_{2}\right)\right\rangle =U^{\alpha}\left(t_{2},t_{1}\right)\left|\Psi^{\alpha}\left(t_{1}\right)\right\rangle$, where $U^{\alpha}\left(t_{2},t_{1}\right)\equiv U^{\alpha}\left(t_{2}^{\alpha},t_{1}^{\alpha}\right)$ has the form \cite{stefanucci_nonequilibrium_2013}
\begin{equation}
U^{\alpha}\left(t_{2},t_{1}\right)=\hat{T}_{C}\exp\left[-\frac{i}{\hbar}\int_{t_{1}^{\alpha}}^{t_{2}^{\alpha}}d\tau H^{\alpha}\left(\tau\right)\right]
\label{eqn:Branch_Propagator}
\end{equation}

and $\hat{T}_{C}$ orders operators chronologically (latest to the left) on $C_{f}$, and anti-chronologically on $C_{b}$.

\section{One Fixed Point}

A sequence of events in time corresponds to a sequence of time-indexed projectors in the consistent histories language, and we now construct a model of an event on the Keldysh contour suitable for combination into similar history sequences. 

We may isolate temporal parts of $\left|\Psi_{U}\right\rangle$ from the main tensor product of Eq. (\ref{eqn:Psi_Universe}). A \textit{fixed point} has identical parts on the two contour branches, corresponding to a `turning point' on the Keldysh contour at time $t_{1}$, i.e. to a point at which the time propagation along $C$ switches from the upper to the lower branch \cite{stefanucci_nonequilibrium_2013}:

\textbf{Definition 2} (Fixed Point) 

\textit{A fixed point at time $t$ is a temporal part of the wavefunction in the $\mathcal{H}_{t}^{b}\otimes\mathcal{H}_{t}^{f}$ subspace, with equal $f$ and $b$ parts.}

Given a specification (via a preparation measurement or theoretical description) of the state $\left|\psi\right\rangle$ of a system at some time $t_{1}$, all quantum histories in $\left|\Psi_{U}\right\rangle$ consistent with this specification are constrained regardless of the contour branch. So there is a fixed point state at $t_{1}$, which is denoted:
\begin{equation}
\left\llbracket \psi\right\rrbracket_{t_{1}} \equiv\left|\psi^{b}\left(t_{1}\right)\right\rangle \otimes\left|\psi^{f}\left(t_{1}\right)\right\rangle
\label{eqn:1FP_Compact}
\end{equation}

\begin{figure}
  \includegraphics[width=.85\linewidth]{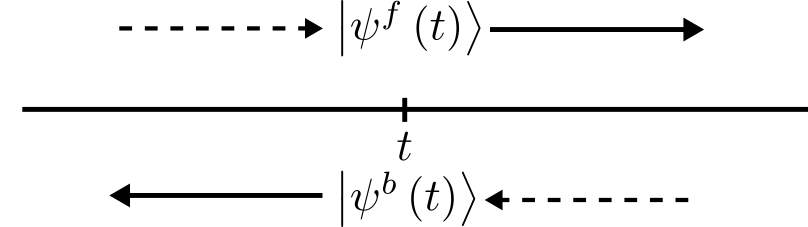}
  \caption{A single fixed point on the Keldysh contour.}
  \label{fig:1FP}
\end{figure}

This corresponds to an event in which the state is specified with definite properties at $t_{1}$ (or a time-indexed projection, in the consistent histories language). We may think of the `present' time $t$ as `pinched' in between the upper-branch and lower-branch times $t^{f}$, $t^{b}$. The fixed point state connects to other points on $C$ in both time directions, in accordance with Eq. (\ref{eqn:Branch_TDSE}). It is represented on $C$ in Fig. \ref{fig:1FP}: the forward-directed part of the fixed point defined at $t$ travels to times occurring `later' than $t^{f}$ on $C_{f}$, and the backward-directed part travels to times occurring `later' than $t^{b}$ on $C_{b}$. Each fixed point is connected to four temporal regions: it acts as a `source' of wavefunction in both time directions (the thick black arrows on Fig. \ref{fig:1FP}), and a `sink' for parts of the wavefunction propagating from times lying `earlier' on $C$ (dashed lines on Fig. \ref{fig:1FP}). Thus, for a full description of a measurement connecting times across the region $\left[t_{1},t_{2}\right]$ at least two fixed points are required, $N_{t}\geq2$ in Eq. (\ref{eqn:Psi_Universe}). A quantum history sequence is defined in these terms:

\textbf{Definition 3} (Quantum history)

\textit{A quantum history $\left|h_{\mathbf{k}}\right\rangle$ extending across the time range $\left[t_{1},t_{2}\right]$ is a product state constructed from a sequence ${\mathbf{k}}=\left\langle k_{1},...,k_{N_{t}}\right\rangle $ of $N_{t}\geq2$ fixed points
\begin{equation}\label{eq:Quantum_History}
    \left|h_{\mathbf{k}}\right\rangle=\overset{N_{t}}{\underset{i=1}{\otimes}}\left\llbracket \psi_{k_{i}}\right\rrbracket _{t_{i}}
\end{equation}
connected by unitary mappings and bounded by fixed points at $t_{1}$ and $t_{2}$.}

In Eq. (\ref{eq:Quantum_History}), each $k_{i}$ in a history $\left|h_{\mathbf{k}}\right\rangle$ ranges over a complete basis set spanning $\mathcal{H}_{t_{i}}^{\alpha}$. To allow us to apply the usual rules of probabilistic reasoning to quantum histories, we define a \emph{family} of quantum histories $\mathcal{F}_{H}$ by imposing the consistency condition that any pair of histories in a family $\left\{ \left|h_{\mathbf{k}}\right\rangle \right\} $ must be non-overlapping:
\begin{equation}\label{eq:Quantum_History_Consistency}
    \left\langle h_{\mathbf{l}}\right.\left|h_{\mathbf{k}}\right\rangle =\delta_{\mathbf{kl}},
\end{equation}

where $\mathbf{k}\neq\mathbf{l}$ if $\left\llbracket \psi_{k_{i}}\right\rrbracket _{t_{i}}\neq\left\llbracket \psi_{l_{i}}\right\rrbracket _{t_{i}}$ for at least one value of $i \in \left[1,...,N_{t}\right]$. Each set of quantum histories provides distinct but complementary descriptions of the system over time, which may or may not correspond to measurement events. Note that the consistency condition Eq. \eqref{eq:Quantum_History_Consistency} prevents the overlap of histories composed of different numbers of times $N_{t}$.

Now, following the terminology of Vaidman \cite{vaidman_schizophrenic_1998}, the \emph{measure of existence} of a history may be defined as the relative size of the wavefunction region occupied by that history.

\textbf{Definition 4} (Measure of existence)

\textit{The measure of existence $m\left(h_{\textbf{k}}\right)$ of a quantum history $\left|h_{\mathbf{k}}\right\rangle$ containing $N_{t}$ fixed points in the time range $\left[t_{1},t_{2}\right]$, is the ratio of the integral of the wavefunction $\triangle\Psi_{\textbf{k}}$ along this history, to that of all histories 
\begin{equation}
    m\left(h_{\textbf{k}}\right)=\frac{\ensuremath{\triangle\Psi_{\textbf{k}}}}{\underset{\mathbf{k'}}{\sum}\ensuremath{\triangle\Psi_{\textbf{k'}}}}
\end{equation}
in a family $\mathcal{F}_{H}$ consistent with the fixed point boundary conditions at $t_{1}$ and $t_{2}$.
}

Fixed point boundary conditions are imposed by taking the inner product of the integrated wavefunction with the `sink' state defined at the upper limits of the $2(N_{t}-1)$ segment integrals. Definition 4 gives precise meaning to the fraction of wavefunction connecting distinct events, and therefore (by Definition 1) a precise foundation for quantum probability:

\begin{tcolorbox}
\textbf{Statistical postulate (Vaidman rule)}
\textit{The quantum probability of a quantum history is equal to its measure of existence in the universal wavefunction.} 
\end{tcolorbox}

Note that no explicit formula has been assumed for the measure of existence. The Vaidman rule is a \emph{conceptual} postulate about the physical foundation of measurement statistics. It remains to be proven that this postulate implies the correct \emph{mathematical} formalism in the case of a quantum measurement.

\section{The Born measure}

By \textbf{Measurement Physicality} and \textbf{Self-Location}, the physical process of an experiment occurs within a region of $\left|\Psi_{U}\right\rangle$ subject to the boundary constraints determined by the preparation. The measure of existence is now evaluated for the simplest type of quantum history - a two-time measurement - with $N_{t}=2$.

Consider a measurement of $\left|\phi\left(t_{2}\right)\right\rangle$ following a preparation of the state $\left|\psi\left(t_{1}\right)\right\rangle$. Without loss of generality, the prepared and measured states are taken to be members of complete bases $\left|\psi\right\rangle = \left|\psi_{1}\right\rangle \in \left\{ \left|\psi_{i}\right\rangle\right\}$, $\left|\phi\right\rangle = \left|\phi_{1}\right\rangle \in \left\{ \left|\phi_{i}\right\rangle\right\}$. The measurement then defines a family of histories 

\begin{equation}
\mathcal{F}_{H}:\left\llbracket \psi\right\rrbracket _{t_{1}}\otimes\left\{ \left\llbracket \phi_{i}\right\rrbracket _{t_{2}}\right\} 
\label{eqn:2FP_Family}
\end{equation}

whose measure of existence in $\left|\Psi_{U}\right\rangle$ can be evaluated. 

The `source' term in this measurement process is the following state constructed from two fixed points (using the notation from Eq. (\ref{eqn:1FP_Compact})):
\begin{equation}
\Psi\left(t_{2}^{b},t_{2}^{f},t_{1}^{b},t_{1}^{f}\right)=\left\llbracket\phi\right\rrbracket_{t_{2}}\otimes\left\llbracket\psi\right\rrbracket_{t_{1}}
\label{eqn:2FPs}
\end{equation}

The total change of this wavefunction across the time interval $t\in\left[t_{1},t_{2}\right]$ is computed by`filling in' the Keldysh contour branches connecting the two fixed points. The exact differential of the wavefunction in Eq. (\ref{eqn:2FPs}) constrained to this region is
\begin{equation}
d\Psi=\frac{\partial\Psi\left(t_{2}^{b},t_{2}^{f},t_{1}^{b},t_{1}^{f}\right)}{\partial t_{1}^{f}}dt_{1}^{f}+\frac{\partial\Psi\left(t_{2}^{b},t_{2}^{f},t_{1}^{b},t_{1}^{f}\right)}{\partial t_{2}^{b}}dt_{2}^{b}
\label{eqn:Exact_Differential_1}
\end{equation}
i.e. one may consider time integrations in both the forwards direction originating at $t_{1}^{f}$ and in the backwards direction from $t_{2}^{b}$. The total change in wavefunction is computed from the line integral along $C$, taking the path $\left(t_{1}^{f},t_{2}^{b}\right)\rightarrow\left(t_{2}^{f},t_{2}^{b}\right)\rightarrow\left(t_{2}^{f},t_{1}^{b}\right)$:
\begin{equation}
D\Psi=\int_{t_{1}^{f}}^{t_{2}^{f}}\frac{\partial\Psi\left(t_{2}^{b},t_{2}^{f},t_{1}^{b},x\right)}{\partial x}dx+\int_{t_{2}^{b}}^{t_{1}^{b}}\frac{\partial\Psi\left(y,t_{2}^{f},t_{1}^{b},t_{2}^{f}\right)}{\partial y}dy
\label{eqn:Line_Integral_1}
\end{equation}

Applying the branch TDSE in Eq. (\ref{eqn:Branch_TDSE}) to the independent degrees of freedom in Eq. (\ref{eqn:2FPs}) and allowing for cancellations, this becomes
\begin{equation}
D\Psi=\left(U^{b}\left(t_{1}^{b},t_{2}^{b}\right)U^{f}\left(t_{2}^{f},t_{1}^{f}\right)-\hat{\mathbf{I}}\right)\Psi\left(t_{2}^{b},t_{2}^{f},t_{1}^{b},t_{1}^{f}\right)
\label{eqn:Wavefunction_Change}
\end{equation}
where $\hat{\mathbf{I}}$ denotes the identity on the Hilbert space $\mathcal{H}_{t_{2}}^{b}\otimes\mathcal{H}_{t_{2}}^{f}\otimes\mathcal{H}_{t_{1}}^{b}\otimes\mathcal{H}_{t_{1}}^{f}$ and the compact notation $U^{b}\left(t_{1}^{b},t_{2}^{b}\right)U^{f}\left(t_{2}^{f},t_{1}^{f}\right) \equiv U^{b}\left(t_{1}^{b},t_{2}^{b}\right)\otimes \hat{\mathbf{I}}_{t_{2}^{f}}\otimes \hat{\mathbf{I}}_{t_{1}^{b}}\otimes U^{f}\left(t_{2}^{f},t_{1}^{f}\right)$ is used. 

\begin{figure}
  \includegraphics[width=\linewidth]{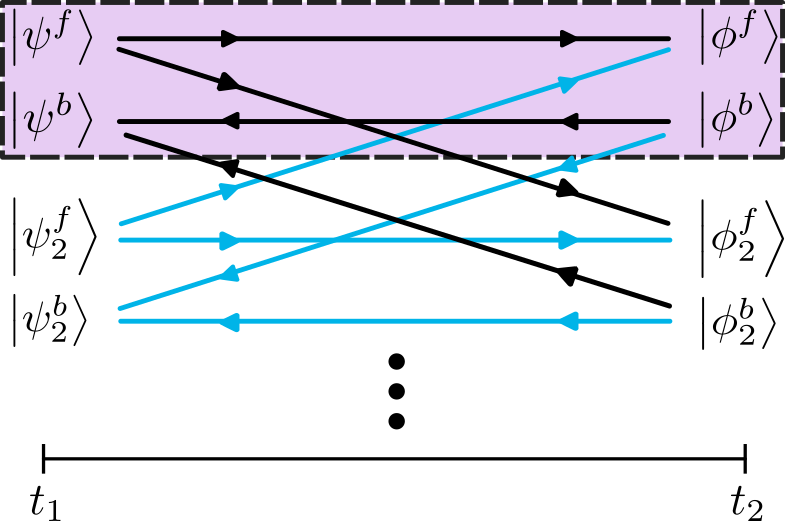}
  \caption{The purple region represents the measure of existence connecting the prepared fixed point state $\left\llbracket\psi\right\rrbracket_{t_{1}}$ to a measurement at $t_{2}$ described by the fixed point $\left\llbracket\phi\right\rrbracket_{t_{2}}$. The black lines represent processes connected to the prepared state. The blue lines represent regions of the universal wavefunction that are incompatible with the preparation.}
  \label{fig:2FPs_Born}
\end{figure}

Unitary evolution from any fixed point produces quantum superpositions represented by a network structure connecting it to other fixed points in the future and past, as shown in Fig. \ref{fig:2FPs_Born}. In this figure, the arrows indicate the temporal orientation of quantum processes. Each fixed point in the expansion of the full wavefunction at a given time is a node where processes begin and terminate in the network, similarly to Ref. \cite{oreshkov_operational_2015}.

The region of wavefunction constrained by two fixed points is represented by the purple region in Fig. \ref{fig:2FPs_Born}. All the processes consistent with the preparation, defining the family $\mathcal{F}_{H}$ are represented by black lines. Blue lines represent those processes not connected to the fixed point state $\left\llbracket\psi\right\rrbracket_{t_{1}}$. From the birds-eye perspective of the universal wavefunction, there is no difference between the black- and blue-line processes. However, they provide a useful distinction for the observer, who thereby determines the region of wavefunction corresponding to their experiment. Formally, if a single node is connected to $N$ others then there will be $N$ Keldysh contour regions and $2N$ separate time branches connected to this node. Moreover, we can view each node as both the source and the sink of all processes connected to it, in the following sense: If a fixed point at time $t$ is connected to $N_{t_{1}}$ nodes at a time $t_{1} < t$, and to $N_{t_{2}}$ nodes at a time $t_{2} > t$, then it is the source of exactly $N_{t_{1}} + N_{t_{2}}$ branch lines which flow away from it, and the sink for the same number of lines which flow into it from other times. If $N_{t_{1}}$ nodes at $t_{1}$ are connected to $N_{t_{2}}$ nodes at $t_{2}$, then there are $2N_{t_{1}}N_{t_{2}}$ branch lines connecting this pair of times, defining $N_{t_{1}}N_{t_{2}}$ regions of the wavefunction, or two-way channels, connecting pairs of fixed points at these times. Every line is in one-to one mapping with a directed process connecting a pair of fixed points in the wavefunction. The amplitude for a process connecting state $\left|\beta\right\rangle$ at time $t_{a}$ to the state $\left|\gamma\right\rangle$ at time $t_{b}$ following propagation along the Keldysh branch $C_{\alpha}$ is given by 

\begin{equation}
c_{\gamma\beta}^{\alpha}\left(t_{b},t_{a}\right)\equiv\left\langle \gamma^{\alpha}\left(t_{b}\right)\right|U^{\alpha}\left(t_{b},t_{a}\right)\left|\beta^{\alpha}\left(t_{a}\right)\right\rangle
\label{eqn:C_amplitude}
\end{equation}

The integrated wavefunction (the purple region in Fig. \ref{fig:2FPs_Born}) connects the two fixed points $\left\llbracket\psi\right\rrbracket_{t_{1}}$ and $\left\llbracket\phi\right\rrbracket_{t_{2}}$, a constraint imposed by taking the inner product of (\ref{eqn:Wavefunction_Change}) with $\left|\psi^{b}\left(t_{1}\right)\right\rangle \left|\phi^{f}\left(t_{2}\right)\right\rangle\left|\psi^{b}\left(t_{1}\right)\right\rangle \left|\phi^{f}\left(t_{2}\right)\right\rangle$ - the `sink' state defined at the upper limits of the integration
\begin{equation}
\triangle\Psi\left[\psi\left(t_{1}\right);\phi\left(t_{2}\right)\right]=c_{\psi\phi}^{b}\left(t_{1},t_{2}\right)c_{\phi\psi}^{f}\left(t_{2},t_{1}\right),
\label{eqn:Wavefunction_Change_Constrained}
\end{equation}
where the overlap with the second term in Eq. (\ref{eqn:Wavefunction_Change}) vanishes since $\left\langle \gamma^{\alpha}\left(t_{b}\right)\right|\left.\beta^{\alpha}\left(t_{a}\right)\right\rangle =0$ given $\left|\beta^{\alpha}\left(t_{a}\right)\right\rangle \in\mathcal{H}_{t_{a}^{\alpha}}$ and $\left|\gamma^{\alpha}\left(t_{b}\right)\right\rangle \in\mathcal{H}_{t_{b}^{\alpha}}$ with $t_{a}^{\alpha}\neq t_{b}^{\alpha}$. Eq. (\ref{eqn:Wavefunction_Change_Constrained}) is just a scalar-valued function, so contour branch labels can be dropped. We now divide by the normalization factor across all measurement outcomes consistent with the preparation, $\sum_{i}\triangle\Psi\left[\psi\left(t_{1}\right);\phi_{i}\left(t_{2}\right)\right]=1$, to give the measure of existence of this history:
\begin{align}\label{eqn:Born_Rule}
m\left(h_{\left\langle \psi,\phi\right\rangle }\right)=
\frac{\triangle\Psi\left[\psi\left(t_{1}\right);\phi\left(t_{2}\right)\right]}{\sum_{i}\triangle\Psi\left[\psi\left(t_{1}\right);\phi_{i}\left(t_{2}\right)\right]}\nonumber\\
=\left|\left\langle \psi\left(t_{1}\right)\right|U\left(t_{1},t_{2}\right)\left|\phi\left(t_{2}\right)\right\rangle \right|^{2}
\end{align}

This gives the relative amount of wavefunction connecting the fixed point $\left\llbracket\psi\right\rrbracket_{t_1}$ to $\left\llbracket\phi\right\rrbracket_{t_2}$ as a proportion of the total region of wavefunction at $t_{2}$ connected to  $\left\llbracket\psi\right\rrbracket_{t_1}$ on the Keldysh contour. 

Eq. (\ref{eqn:Born_Rule}) is the core result of this work: the measure of existence of a quantum history describing a quantum measurement process equals the Born measure. The power of 2 in this measure is a direct result of the two time branches in $C$. Instead of postulating the mathematical form of the measure of existence \cite{vaidman_derivations_2020}, the Born measure has been derived from the temporal structure of the multiple-Keldysh-time wavefunction and the Vaidman rule.

\section{Three Fixed Points}

The ABL rule may be derived from the Born measure \cite{aharonov_two-state_2008}, but will now be derived as the measure of existence of the quantum history connecting $N_{t}=3$ fixed points. This case is important to consider because it involves all four regions of the Keldysh contour connected to the intermediate fixed point. 

\begin{figure}
  \includegraphics[width=\linewidth]{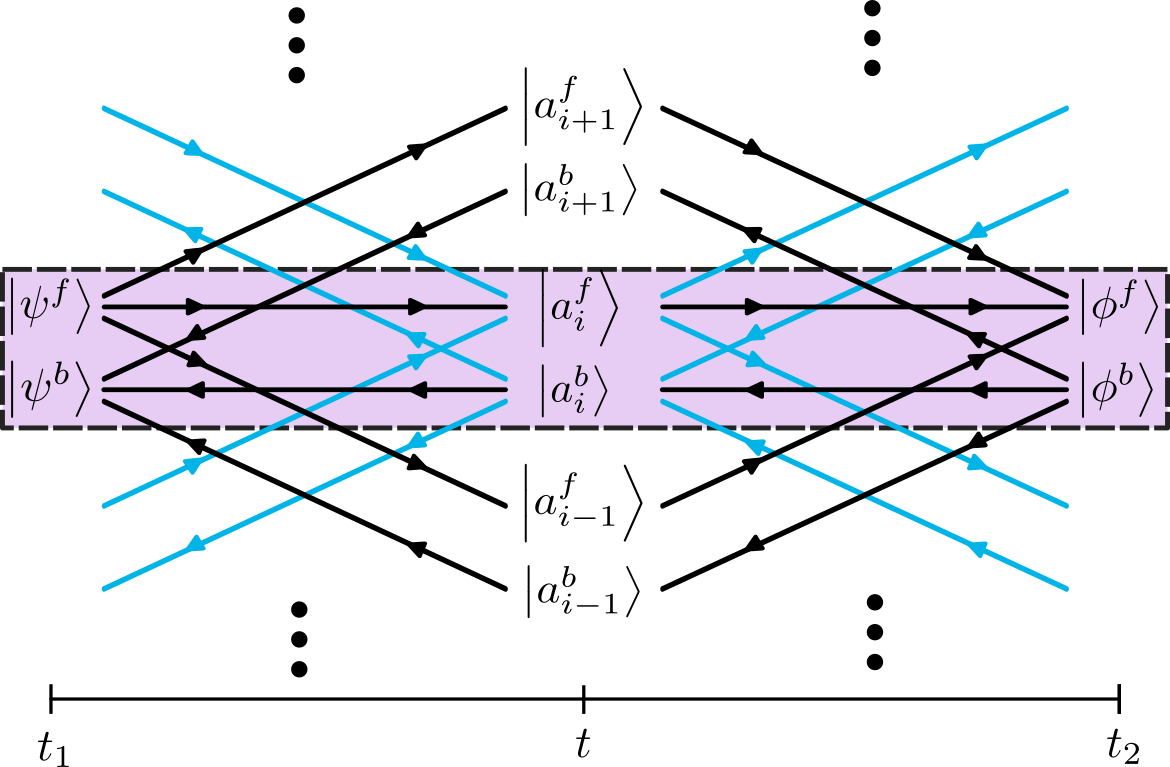}
  \caption{The purple region represents the measure of existence corresponding to the ABL measure in an experiment connecting the pre- and post-selection fixed points $\left\llbracket\psi\right\rrbracket_{t_{1}}$ and $\left\llbracket\phi\right\rrbracket_{t_{2}}$ to a measurement at $t$ corresponding to the fixed point $\left\llbracket a_{i} \right\rrbracket_{t}$.}
  \label{fig:FPF_ABL}
\end{figure}

Suppose that pre- and post-selection measurements at $t_{1}$ and $t_{2}$ yield the states $\left|\psi\right\rangle$ and $\left|\phi\right\rangle$, respectively. One is then interested in the probability of measuring a state in some basis, $\left| a_{i} \right\rangle \in \left\{ \left|a_{k}\right\rangle\right\}$, at the measurement time $t$, where $t_{1}<t<t_{2}$. This experiment corresponds to the family of histories

\begin{equation}
    \mathcal{F}_{H}:\left\llbracket \psi\right\rrbracket _{t_{1}}\otimes\left\{ \left\llbracket a_{i}\right\rrbracket _{t}\right\} \otimes\left\llbracket \phi\right\rrbracket _{t_{2}} \label{3FP_family}   
\end{equation}

as represented schematically in Figure \ref{fig:FPF_ABL}, where the purple region represents the measure of existence corresponding to the measurement of $\left| a_{i} \right\rangle$ at time $t$, black lines define the history family $\mathcal{F}_{H}$ (processes consistent with the pre- and post-selected boundary values), and blue lines represent wavefunction regions that are incompatible with the preparation. 

This experimental situation is described by the following `source' wavefunction constructed from three fixed points:
\begin{equation}
\Psi\left(t_{2}^{b},t_{2}^{f},t^{b},t^{f},t_{1}^{b},t_{1}^{f}\right)=\left\llbracket\phi\right\rrbracket_{t_{2}}\otimes\left\llbracket a_{i}\right\rrbracket_{t}\otimes\left\llbracket\psi\right\rrbracket_{t_{1}}
\label{eqn:3FPs}
\end{equation}

\begin{figure}[htp]

  \begin{subfigure}{\linewidth}
    \subcaption{FPF} \label{fig:5a}
    \includegraphics[clip, width=\linewidth]{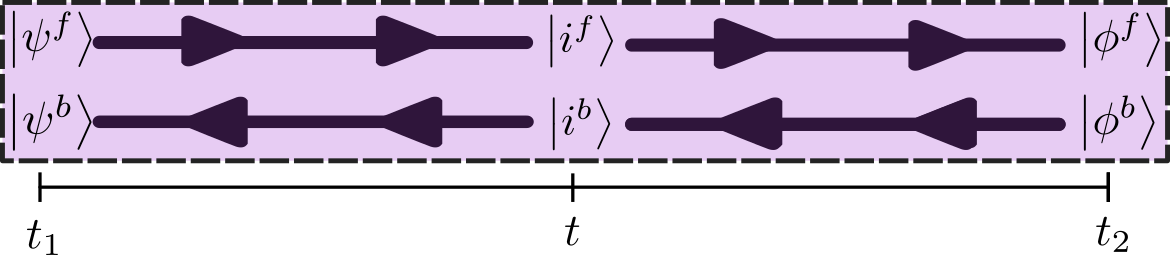}
  \end{subfigure}%

  \begin{subfigure}{\linewidth}
    \subcaption{TSVF} \label{fig:5b}
    \includegraphics[clip, width=\linewidth]{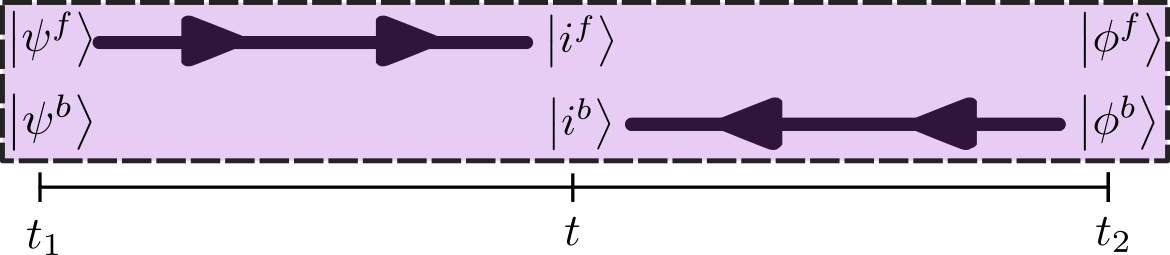}
  \end{subfigure}%
  
  \begin{subfigure}{\linewidth}
    \subcaption{Schr{\"o}dinger} \label{fig:5c}
    \includegraphics[clip, width=\linewidth]{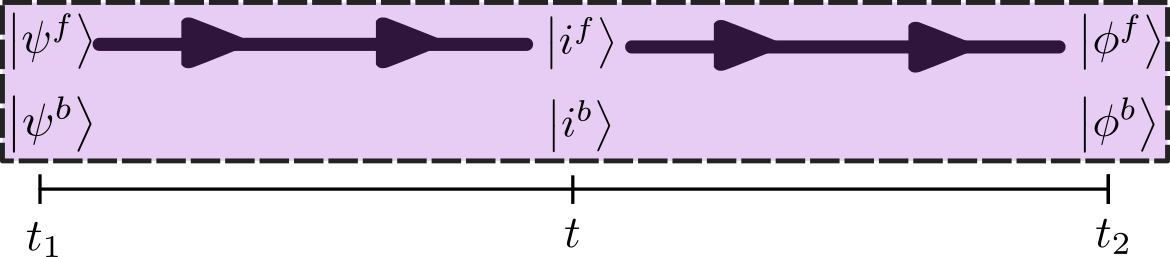}
  \end{subfigure}
\caption{Schematic representation of the regions and direction of time propagation between three consecutive boundary conditions considered within (a) the FPF, (b) the TSVF and (c) standard Schr{\"o}dinger dynamics.} \label{fig:5}
\end{figure}

The fixed points in this state are connected via black lines in the shaded region of Fig. \ref{fig:FPF_ABL}. By contrast, the TSVF divides the universe into `future' times, described by a state vector travelling backwards from the post-selection, and `past' times, described by future-oriented propagation from the pre-selection. This restricts the dynamics to the upper black arrow left of the measurement time ($<_{C}t^{f}$) and the lower black arrow right of the measurement time ($<_{C}t^{b}$) on Fig. \ref{fig:FPF_ABL}, effectively throwing away half of the wavefunction by treating the fixed point at $t$ as a sink only. The propagation between three boundary constraints in the FPF is illustrated schematically in Fig. \ref{fig:5} (a), where all sections of the contour are covered. This is compared to the situation in the TSVF in \ref{fig:5} (b), where only half of the available contour is included, therefore violating \textbf{Event Symmetry}. For comparison, the standard Schr{\"o}dinger dynamics used in the consistent histories framework is illustrated in \ref{fig:5} (c). We also note that the TSVF formalism allows oppositely-oriented states to overlap at the intermediate measurement time (the backwards-travelling vector from the future is represented as a `bra' state in the conjugate Hilbert space $\mathcal{H}_{t_{2}}^{\dagger}$) \cite{aharonov_two-state_2008}, which is prevented by branch-independence in the FPF.

The total wavefunction along segments of the Keldysh contour connecting the three fixed points in Eq. (\ref{eqn:3FPs}) is a line integral of the exact differential 
\begin{equation}
d\Psi=\frac{\partial\Psi}{\partial t_{1}^{f}}dt_{1}^{f}+\frac{\partial\Psi}{\partial t^{f}}dt^{f}+\frac{\partial\Psi}{\partial t_{2}^{b}}dt_{2}^{b}+\frac{\partial\Psi}{\partial t^{b}}dt^{b}
\label{eqn:Exact_Differential_2}
\end{equation}

along the path $\left(t_{2}^{b},t^{b},t^{f},t_{1}^{f}\right)\rightarrow\left(t_{2}^{b},t^{b},t^{f},t^{f}\right)\rightarrow\left(t_{2}^{b},t^{b},t_{2}^{f},t^{f}\right)\rightarrow\left(t^{b},t^{b},t_{2}^{f},t^{f}\right)\rightarrow\left(t^{b},t_{1}^{b},t_{2}^{f},t^{f}\right)$, which is the integral path along the horizontal black lines enclosed by the purple shading in Fig. \ref{fig:FPF_ABL}. Since the four time degrees of freedom are independent, the total wavefunction in the temporal region $t\in\left[t_{1},t_{2}\right]$ is
\begin{gather} \label{3FPS_Change}
 D\Psi=U^{b}\left(t^{b},t_{2}^{b}\right)\left\llbracket\phi\right\rrbracket_{t_{2}} \\
 \otimes U^{b}\left(t_{1}^{b},t^{b}\right)U^{f}\left(t_{2}^{f},t^{f}\right)\left\llbracket a_{i} \right\rrbracket_{t}
 \otimes U^{f}\left(t^{f},t_{1}^{f}\right)\left\llbracket\psi\right\rrbracket_{t_{1}}-\Psi\nonumber
\end{gather}
Taking the inner product of $D\Psi$ with the corresponding `sink state' $\left|a_{i}^{b}\left(t\right)\right\rangle \left|\phi^{f}\left(t_{2}\right)\right\rangle \left|\psi^{b}\left(t_{1}\right)\right\rangle \left|\phi^{f}\left(t_{2}\right)\right\rangle \left|\psi^{b}\left(t_{1}\right)\right\rangle \left|a_{i}^{f}\left(t\right)\right\rangle$ and then normalizing gives the measure of existence of a history connecting the fixed points $\left\llbracket\psi\right\rrbracket_{t_{1}}$, $\left\llbracket a_{i} \right\rrbracket_{t}$ and $\left\llbracket\phi\right\rrbracket_{t_{2}}$(the region covered by black lines in Fig. \ref{fig:FPF_ABL}):
\begin{gather}\label{eqn:ABL_Rule}m\left(h_{\left\langle \psi,a_{i},\phi\right\rangle }\right)=
\frac{\triangle\Psi\left[\psi\left(t_{1}\right);a_{i}\left(t\right);\phi\left(t_{2}\right)\right]}{\underset{k}{\sum}\triangle\Psi\left[\psi\left(t_{1}\right);a_{k}\left(t\right);\phi\left(t_{2}\right)\right]}\\
=\frac{\left|\left\langle \phi\left(t_{2}\right)\right|U\left(t_{2},t\right)\left|a_{i}\left(t\right)\right\rangle \left\langle a_{i}\left(t\right)\right|U\left(t,t_{1}\right)\left|\psi\left(t_{1}\right)\right\rangle \right|^{2}}{\underset{k}{\sum}\left|\left\langle \phi\left(t_{2}\right)\right|U\left(t_{2},t\right)\left|a_{k}\left(t\right)\right\rangle \left\langle a_{k}\left(t\right)\right|U\left(t,t_{1}\right)\left|\psi\left(t_{1}\right)\right\rangle \right|^{2}}\nonumber
\end{gather}

Thus we recover the ABL rule. It has been derived as a ratio of wavefunction regions integrated over $C$. There is no stochastic `collapse' process, just unitary evolution and the imposition of constraints.. 

This analysis is easily extended to a sequence of measurements - each fixed point increases the dimensionality of the line integral in Eq. (\ref{eqn:Line_Integral_1}) by two, so the corresponding change $\triangle\Psi$ in a $N_{t}$-time history is obtained from the $2(N_{t}-1)$-dimensional line integral along the relevant Keldysh contour segments.

\section{Discussion}

In this paper we have derived a direct connection between the temporal structure of the wavefunction and the Born rule of quantum mechanics. Central to our thesis is the concept of a ‘fixed point’, which replaces the initial condition of standard quantum theory with a state that serves as both ‘source’ and ‘sink’ in both directions of time, defined on the Keldysh contour. The FPF has many advantages:

\begin{itemize}
    \item Unlike derivations which appeal to contingent initial or final conditions of the universe \cite{gell-mann_quantum_1996,durr_bohmian_2004,aharonov_two-time_2017}, it explains the ubiquity of the Born measure in nature from temporally local constraints.
    \item It does not contradict the thermodynamic arrow of time: the forwards and backwards oriented parts of the wave function combine to produce observed reality. Indeed, directional processes such as dissipation and relaxation to nonequilibrium steady states in open quantum systems are described by Keldysh-based methods on a routine basis \cite{ridley_fluctuating-bias_2016,tuovinen_adiabatic_2019,tuovinen_comparing_2020,ridley_quantum_2021}.  
    \item It describes deterministic unitary quantum mechanics with a multiple-event structure which may have implications for quantum gravity \cite{maccone_fundamental_2019}, as it makes no fundamental distinction between past and future times - a fixed point is simply a crossing point for quantum histories. In this framework, the statistical postulate supplies the \textit{meaning} of probability, and ontic structure supplies its \textit{mathematical form}.
    \item It is logically simpler than approaches to quantum probability which involve deviations from unitarity \cite{ghirardi_unified_1986,vinante_improved_2017} or the introduction of additional ontology \cite{bohm_suggested_1952,durr_bohmian_2004}.
    \item It contains no genuine randomness, only integrals over temporal regions of the wavefunction, which is fixed in time.
\end{itemize}

Hitherto, Zurek's `envariance'-based approach to quantum probabilities was the leading candidate for a physical derivation \cite{zurek_environment-assisted_2003,zurek_probabilities_2005}. This strategy relies upon (i) the Schmidt decomposition into entangled system + environment states via decoherence, (ii) `envariance' symmetry-based probability assignments and (iii) the modification of the environment with ancilla states satisfying certain `fine-graining' properties. By contrast, the argument in this paper (i) assumes nothing about the internal composition of states beyond the ontological and dynamical postulates (ii) assumes nothing about probabilities beyond the statistical postulate and (iii) has no dependence on details of the environment.

Since we are here considering unitary wave mechanics only, the FPF supports an Everettian interpretation of the quantum theory \cite{everett_relative_1957} with the caveat that branching of the wavefunction is permitted in both time directions. The other candidate for a time-symmetric quantum theory considered here - the TSVF - omits crucial information contained in the full Keldysh time structure. It is this temporal structure which explains the emergence of quantum probability.

\begin{acknowledgements} 
I am grateful to Yakir Aharonov, Alessandro de Vita, Lev Kantorovich, Don Page, Riku Tuovinen and Lev Vaidman for many helpful discussions.
This work has been supported in part by the Israel Science Foundation Grant No. 2064/19 and the National Science Foundation–US-Israel Binational Science Foundation Grant No. 735/18.
\end{acknowledgements}


%

\end{document}